\documentclass[
aps,
pop,
superscriptaddress, 
twocolumn
]{revtex4-2}

\usepackage{graphicx}
\usepackage{dcolumn}
\usepackage{bm}

\usepackage[utf8]{inputenc}
\usepackage[T1]{fontenc}
\usepackage{mathptmx}
\usepackage{etoolbox}
\usepackage{siunitx}
\usepackage{amsmath}
\usepackage{amssymb}

\graphicspath{ {./figs/} }
\begin{document}
\title{Investigating radiatively driven, magnetised plasmas with a university scale pulsed-power generator}
\author{Jack W. D. Halliday}
	\email{jack.halliday12@imperial.ac.uk}
	\address{Blackett Laboratory, Imperial College London, SW7 2AZ, United Kingdom}
\author{Aidan Crilly}
	\address{Blackett Laboratory, Imperial College London, SW7 2AZ, United Kingdom}
\author{Jeremy Chittenden}
	\address{Blackett Laboratory, Imperial College London, SW7 2AZ, United Kingdom}
\author{Roberto C. Mancini}
	\address{University of Nevada, Reno, NV 89557, USA}
\author{Stefano Merlini}
	\address{Blackett Laboratory, Imperial College London, SW7 2AZ, United Kingdom}
\author{Steven Rose}
	\address{Blackett Laboratory, Imperial College London, SW7 2AZ, United Kingdom}
	\address{Clarendon Laboratory, University of Oxford, OX1 3PU, United Kingdom}
\author{Danny R. Russell}
	\address{Blackett Laboratory, Imperial College London, SW7 2AZ, United Kingdom}
\author{Lee G. Suttle}
	\address{Blackett Laboratory, Imperial College London, SW7 2AZ, United Kingdom}
\author{Vicente Valenzuela-Villaseca}
	\address{Blackett Laboratory, Imperial College London, SW7 2AZ, United Kingdom}
\author{Simon N. Bland}
	\address{Blackett Laboratory, Imperial College London, SW7 2AZ, United Kingdom}
\author{Sergey V. Lebedev}
	\address{Blackett Laboratory, Imperial College London, SW7 2AZ, United Kingdom}

\date{\today}

\begin{abstract}
	We present first results from a novel experimental platform which is able to access physics relevant to topics including indirect-drive magnetised ICF; laser energy deposition; various topics in atomic physics; and laboratory astrophysics (for example the penetration of B-fields into HED plasmas). This platform uses the X-Rays from a wire array Z-Pinch to irradiate a silicon target, producing an outflow of ablated plasma. The ablated plasma expands into ambient, dynamically significant B-fields ($\sim \SI{5}{\tesla}$) which are supported by the current flowing through the Z-Pinch. The outflows have a well-defined (quasi-1D) morphology, enabling the study of fundamental processes typically only available in more complex, integrated schemes. Experiments were fielded on the MAGPIE pulsed-power generator (\SI{1.4}{\mega\ampere}, \SI{240}{\nano\second} rise time). On this machine a wire array Z-Pinch produces an X-Ray pulse carrying a total energy of $\sim \SI{15}{\kilo\joule}$ over $\sim \SI{30}{\nano\second}$. This equates to an average brightness temperature of around $\SI{10}{\electronvolt}$ on-target.        
\end{abstract}
\maketitle

\section{\label{sec:Intro}Introduction}
\begin{figure}
	\centering
	\includegraphics[width=\columnwidth]{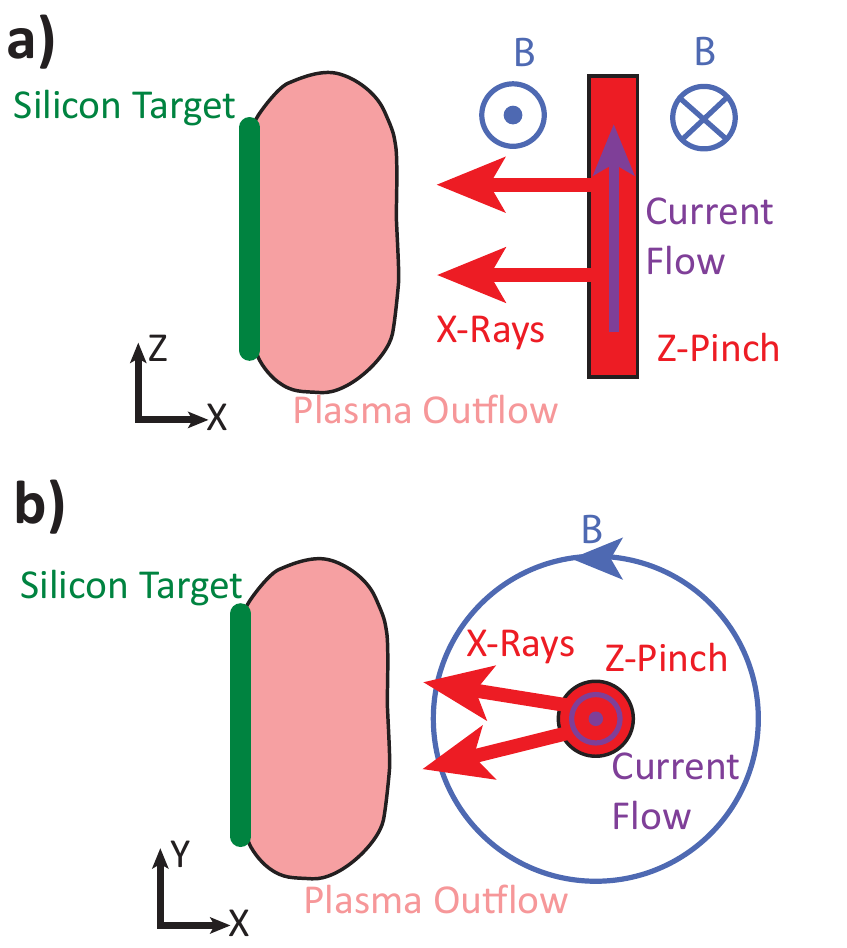}
	\caption{
		Orthogonal views of the experimental setup. These show the position of the wire-array Z-Pinch; the orientation of the magnetic field; and the position of the planar silicon target. The separation between the target and the pinch ranged between $1-\SI{4}{\centi\meter}$. a) Shows a \emph{`side-on'} view of the experiment, where structure is resolved in the $X-Z$ plane. b) Shows an \emph{`end-on'} view of the experiment, where structure is resolved in the $X-Y$ plane.        
	}
	\label{fig:setup}
\end{figure}
In this paper we describe a novel platform (shown in figure \ref{fig:setup}) which uses the radiation pulse from a wire array Z-Pinch to drive plasma ablation from solid silicon targets. The ablated silicon plasmas have a uniform (quasi-1D) structure and expand into an ambient magnetic field produced by the current flowing through the Z-Pinch. 

The platform produces plasma flows with a simple experimental morphology that have well characterised boundary and initial conditions. This means that the platform is a promising test bed for the study of fundamental physics relevant to a wide variety of different topics within the high energy density (HED) plasma physics community, including X-Ray driven ablation physics \cite{Saillard2010}; magnetic field penetration / anomalous diffusion \cite{Brenning2009}; various topics in atomic and radiation physics; and laser energy deposition \cite{Skupsky1987}. 

Because of the simple morphology, the effects of different physical processes can be isolated and studied separately. This is unusual for a HED experiment: Typically much of the physics we describe in this paper is only accessible in more complex integrated schemes. For example, magnetic field diffusion is a physical process which may be well diagnosed using the setup we describe here. This is a process which is also relevant to MAGLIF \cite{Slutz2010, Gomez2020, Velikovich2014} and magnetised-ICF implosions \cite{Davies2017}. However, in these integrated fusion experiments, plasma diagnosis is more challenging and many interacting processes mean that it is challenging to isolate the influence of individual aspects of the physics.          

The remainder of the discussion in this paper is organised as follows: 

\emph{Section \ref{sec:setup} -- }we provide a detailed description of the experimental setup, including quantitative detail on the driving radiation pulse. 

\emph{Section \ref{sec:inter} -- }we present interferometry and self-emission imaging data. These  results show that the ablated silicon plasma has a quasi-1D morphology and that the experimental dynamics are sensitive to the time-history of the driving radiation pulse.

\emph{Section \ref{sec:sims} -- }we describe results from radiative-magnetohydrodynamics simulations of the experiment performed with the code Chimera \cite{Chittenden2016, McGlinchey2018} which show that the presence of an ambient magnetic field affects the electron density profile in the ablated silicon plasma. 

\emph{Section \ref{sec:thomson} -- }we present temperature, velocity, and ionisation measurements from an optical Thomson scattering diagnostic. These results are in qualitative agreement with the simulations and show evidence of Thomson probe heating in the cold, dense plasma near to the target. The results also suggest that the radiation field from the Z-Pinch may play a role in determining the charge state distribution of the ablated silicon plasma.  

\emph{Section \ref{sec:absorbtion} --} we compare results from a shadowgraphy diagnostic to predictions from a theory for inverse Bremsstrahlung absorption and find that the measured absorption coefficient is systematically higher than the predictions from theory.     

\emph{Section \ref{sec:conclusion}} -- we summarise our conclusions.

\section{\label{sec:setup}Experimental setup}
Experiments were performed on the MAGPIE pulsed power generator \cite{Mitchell1996} at Imperial College London (\SI{1.4}{\mega\ampere} peak current, \SI{240}{\nano\second} rise time). This machine was used to drive wire arrays consisting of 32 aluminium wires (\SI{12.7}{\micro\meter} wire thickness, \SI{16}{\milli\meter} array diameter). This load delivered an X-Ray pulse that carried $\sim\SI{15}{\kilo\joule}$ of energy in a $\sim\SI{30}{\nano\second}$ pulse. The initial choice of array parameters represents a compromise between maximising total X-Ray yield and minimising experimental complexity. In future experiments we may consider using different wire materials to change the spectral character of the X-Ray pulse -- this would likely have a small (but not totally negligible) impact on both the yield and pulse duration. By adjusting the array diameter, the wire material, and the wire diameter it would also be possible to control the X-Ray pulse duration / shape.  This would potentially come at some cost to the total X-Ray yield produced in experiments. In principle, it may also be feasible to field a nested wire array: Experiments on the \SI{20}{\mega\ampere} Z-machine have shown that nested arrays yield a \SI{40}{\percent} increase in X-Ray power over a single array \cite{Deeney1998}. It is also possible to control the X-Ray pulse shape through the use of nested wire arrays \cite{Bland2003, Cuneo2005}.    

The separation between the axis of the pinch and silicon targets was varied between \SI{1.5}{\centi\meter} and \SI{4}{\centi\meter}.The silicon target was placed inside of the wire array's return structure and so the ablated silicon plasma  expanded into an ambient magnetic field supported by the current flowing through the Z-Pinch. For a target separation of \SI{4}{\centi\meter}, the peak magnetic field strength was \SI{7}{\tesla}. In future experiments, by reducing the array diameter to \SI{8}{\milli\meter} and the separation between the target and the axis to \SI{6}{\milli\meter}, it would be possible to increase the peak magnetic field strength to as much as \SI{50}{\tesla}.

Figure \ref{fig:setup} is a pair of diagrams depicting this experimental setup from different fields of view. The figure defines the coordinate system which we will use to describe the experiment throughout this paper: The $X$ axis of this coordinate system is aligned with the target normal; the $Z$ axis is along the current path; and the $Y$ axis is directed in the same direction as the magnetic field in the vicinity of the target. Figure \ref{fig:setup}a shows a `\emph{side on}' view of the setup, where structure is resolved in the $X-Z$ plane. Figure \ref{fig:setup}b shows an `\emph{end on}' view of the setup in which structure is resolved in the $X-Y$ plane.              

\begin{figure}
	\centering\includegraphics[width=\columnwidth]{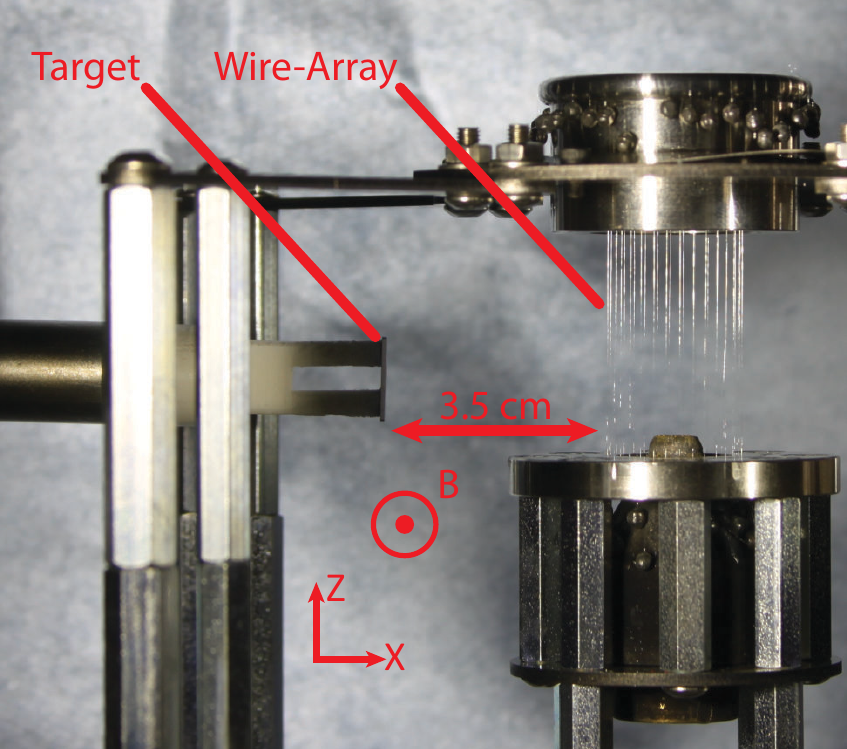}
	\caption{
		A setup photograph of an X-Ray driven silicon ablation experiment. The silicon target (mounted on a two pronged fork and facing the array) sits towards the left hand side of the image, and the wires of the array are on the right hand side of the image. 
	}
	\label{fig:setup-pic}
\end{figure}
A photograph of the experimental setup is shown in figure \ref{fig:setup-pic}. Labels in the figure indicate the position of the silicon target and the initial position of the wires. During an experiment, the generator's current pulse is directed through the wires in the array, resistively heating them to form a plasma. The  current also produces a global, azimuthal magnetic field and so the plasma experiences a $\mathbf{J\times B}$ force, driving it towards the central axis. The plasma stagnates on axis to form a hot, dense column. It is the emission from this stagnated plasma which is the source of the driving X-Rays used in this investigation.

The dynamics of, and emission spectra produced by, wire arrays driven on mega-ampere class pulsed power generators is a topic which has been well described by a number of previous publications (see for example \cite{Bland2007, Shelkovenko2004, Lebedev2002, Lebedev2005, Lebedev2001, Chittenden2004, McBride2009}). The overview we give here is limited to a phenomenological description of wire array emission spectra. This is included to provide context for the discussion of silicon ablation experiments which is the main topic of this paper.      

\begin{figure}
	\centering
	\includegraphics[width=\columnwidth]{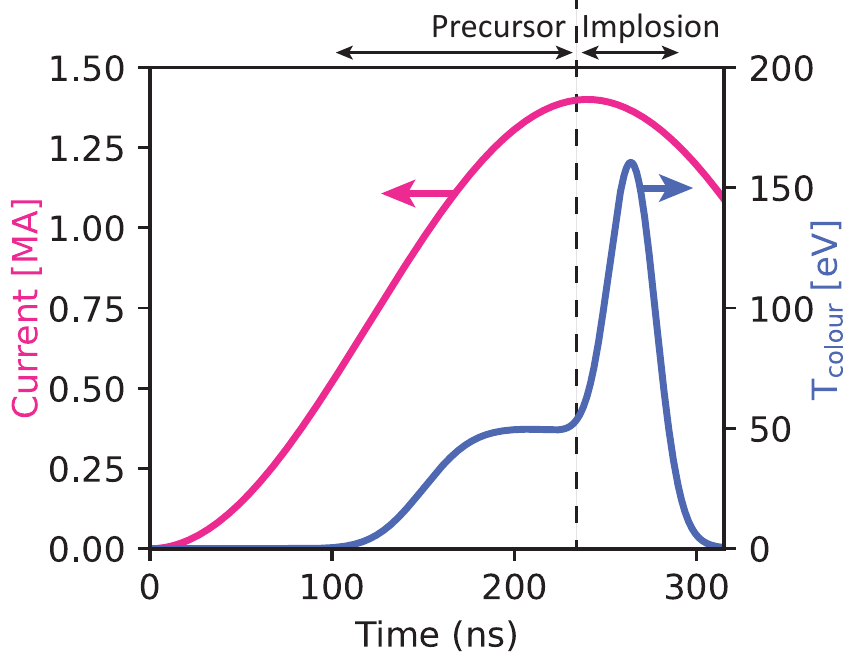}
	\caption{
		A plot showing the variation in the spectral character of the emission from a wire array Z-Pinch as a function of time, overlaid against the generator current pulse. Note that, in reality, the emission from the array is non-thermal and the colour temperatures are purely indicative.
	}
	\label{fig:ideal_fields}
\end{figure}
A plot of a generator current pulse and effective colour temperature as a function of time for a mega-ampere class Z-Pinch implosion is shown in figure \ref{fig:ideal_fields}. Here, colour temperature is defined as the temperature of of a black-body which roughly approximates the spectral character of a non-black-body source. \cite{2009a}. We emphasise that this plot represents a empirical model of the radiative dynamics of an array, informed by experimental data and the results of simulations. The figure shows that, around \SI{100}{\nano\second} after the start of the driving current pulse, the wire array starts to emit so called `precursor' (or pre-pulse) radiation. This radiation is relatively soft (nominal colour temperature $\sim \SI{50}{\electronvolt}$), and radiates roughly \SI{300}{\joule} in total \cite{Bott2006}. The source of precursor radiation is emission from the wires in the array, and from a column of ablated aluminium plasma, created by the (current-driven) ablation of material from the wires, and accelerated towards the axis by the $\mathbf{J \times B}$ force \cite{Lebedev2005}.   

The precursor persists for a timescale $\sim \SI{150}{\nano\second}$ prior to the main implosion of the array. This main implosion happens over a shorter period ($\sim \SI{30}{\nano\second}$) and has a harder spectral character (indicative colour temperature $\sim \SI{150}{\electronvolt}$) \cite{Lebedev2005}. This main implosion phase occurs when approximately half the mass of the wires is lost to ablation, and the wires implode in bulk to produce a hotter, denser X-Ray radiator on the central axis of the array \cite{Lebedev2001, Lebedev2005}.  Bolometery measurements suggest that the total energy radiated in the main implosion is around $10-\SI{15}{\kilo\joule}$ \cite{Bland2007}. For the geometry of the Z-Pinch, this corresponds to a brightness temperature of $\sim \SI{50}{\electronvolt}$ at the emitting surface.

Here, the brightness temperature (sometimes called the effective temperature) of a non-black-body source is defined by the temperature of a black-body with the same (spectrally integrated) intensity. In determining the brightness temperature quoted above we assumed a pinch diameter of \SI{2}{\milli\meter} (indicated by diagnostic images); a pinch length of \SI{20}{\milli\meter} (from the initial conditions); and an emission duration of \SI{30}{\nano\second} (indicated by X-Ray diode measurements).      

Calculations were performed with the view-factor code VISRAD \cite{MacFarlane2003} to investigate how geometric dilution varied the brightness temperature on-target as a function of the separation between the target and the pinch. These results suggest that the brightness temperature on-target ranged from \SI{30}{\electronvolt} (at a separation of \SI{1.5}{\centi\meter}) to \SI{10}{\electronvolt} (at a separation of \SI{4}{\centi\meter}). Due to the relatively large source size, the targets were uniformly illuminated: For a separation of \SI{1.5}{\centi\meter}, the spatial variation of flux on-target seen in VISRAD simulations was less than \SI{5}{\percent}.       

\begin{figure}
	\centering
	\includegraphics[width=\columnwidth]{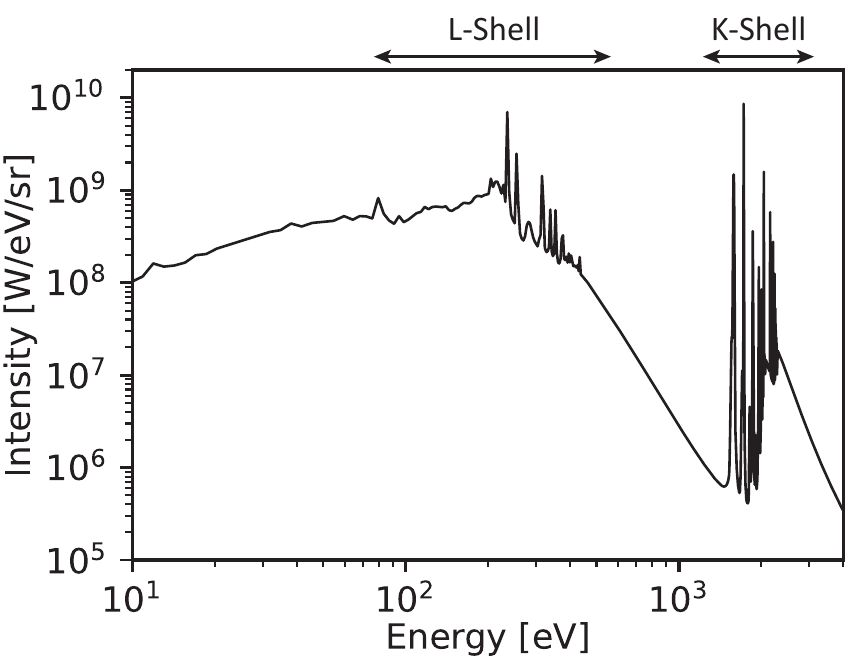}
	\caption{
		A synthetic X-Ray spectrum for an aluminium wire array Z-Pinch, calculated by post processing a Gorgon-MHD \cite{Chittenden2004} simulation using the atomic code SpK \cite{Niasse2011}. The positions of the aluminium K and L shells are also indicated.
	}
	\label{fig:spk_spec}
\end{figure}

An important caveat to our discussion above is that the radiation from the wire array is not completely thermal (see for example the spectra in \cite{Bland2007, McBride2009}). This is demonstrated in figure \ref{fig:spk_spec} which shows a synthetic X-Ray spectrum, calculated by post-processing a simulation of a  Gorgon magnetohydrodynamics (MHD) \cite{Chittenden2004} simulation of a wire array using the SpK atomic code \cite{Niasse2011}. The data shown in the plot is for the spatially integrated emission from the array at the time of peak X-Ray production. It shows that the emission is dominated by bound-bound features in the aluminium L and K shells. Particularly in the L-shell, this forest of emission lines blend together to form a quasi-continuum. Emission from the L-shell is present both for the precursor and the main implosion. Emission from the K-shell is only present during the main implosion. 

The simulations we describe in section \ref{sec:sims} suggest that the bulk of the energy deposited into the silicon originated from L-shell emission features. For photons in this energy range, the penetration depth in solid silicon is on the order of $\sim \SI{100}{\nano\meter}$. 

For the sake of the discussion in this paper, the key details to take away here are that the bulk of the energy deposited into the silicon is from emission during the implosion phase of the wire array. That said, radiation in the precursor phase plays an important role in determining the initial behaviour of the ablated silicon.  

\section{\label{sec:inter}Electron density measurements and experimental dynamics}
In this section we present experimental results from laser interferometry and optical self-emission imaging. The data provides insight into profiles of electron density and qualitative experimental dynamics. 
 
\begin{figure}
	\centering
	\includegraphics[width=\columnwidth]{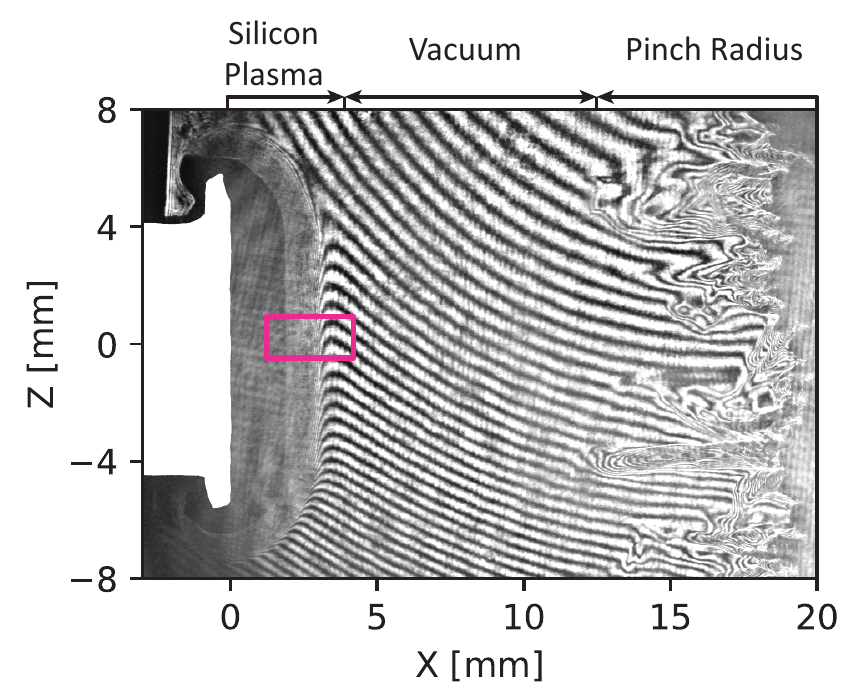}
	\caption{An interferogram of an ablating silicon target (side-on / $X-Z$ plane view), captured \SI{330}{\nano\second} after current start. In the image we see plasma ablated from the target in the region $0 < X < \SI{4}{\milli\meter}$, and plasma from the wire array Z-Pinch in the region $12 < X < \SI{20}{\milli\meter}$. The pink box, centred on $Z=\SI{0}{\milli\meter}$, indicates the region of the interferogram which is shown in figure \ref{fig:density}.
	}
	\label{fig:interferogram}
\end{figure}

A typical laser interferogram (\SI{532}{\nano\meter} probe wavelength, \SI{200}{\femto\second} pulse width), obtained in an experiment with a separation of \SI{20}{\milli\meter} between the silicon target and the Z-Pinch, is shown in figure \ref{fig:interferogram}. For this experiment, the Z-Pinch was configured so the implosion occurred at the time when the generator's current pulse was at a maximum. In the image, the displacement of interference fringes from their position in the absence of plasma is proportional to the line integrated electron density. The interferogram shows  the initial position of the target as a white mask (behind $X=\SI{0}{\milli\meter}$). In the region between $X=\SI{0}{\milli\meter}$ and \SI{2}{\milli\meter}, the plasma is dense enough to absorb the probing laser beam and so the interference fringes are lost. In the region $X = 2 - \SI{4}{\milli\meter}$, fringes are observed and so a measurement of electron density can be made. The signature of the stagnated wire array Z-Pinch (our source of driving X-Rays) can also be seen and is centred on $X=\SI{20}{\milli\meter}$. The error on this measurement of areal density is on the order of one-quarter fringe shift or $1\times 10^{17} \; \si{\per\centi\meter\squared}$.  

From this interferometry data, a striking finding is that the silicon's density profile is constant along the $Z$ direction throughout most of the outflow. In some experiments, simultaneous laser probing was carried out in the orthogonal direction (probing along $Z$ with resolution in the $X-Y$ plane). This revealed a similar degree of uniformity in the $Y$ direction -- suggesting a quasi-1D profile of expanding plasma.                

\begin{figure}
	\centering
	\includegraphics[width=\columnwidth]{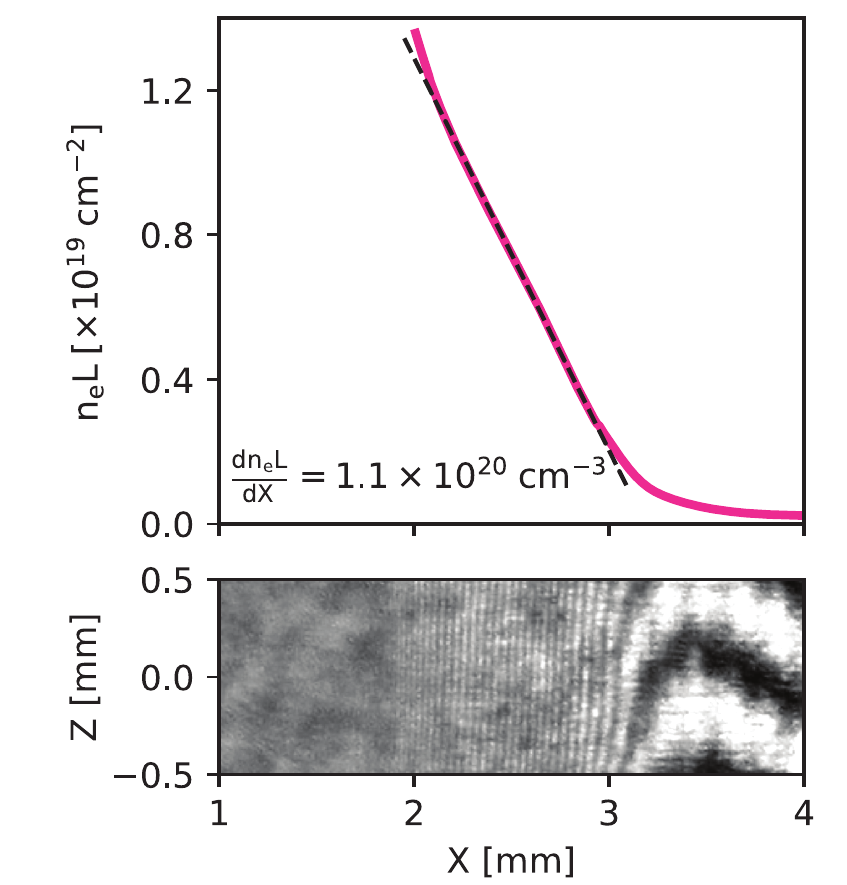}
	\caption{\emph{Top: }Measured profile of line integrated electron density, obtained from the interferogram shown in figure \ref{fig:interferogram}. The data is spatially resolved in $X$, and is averaged over a \SI{1}{\milli\meter} region, centred on $Z=\SI{0}{\milli\meter}$. \emph{Bottom: }Zoomed-in view of the region of the interferogram which was processed to obtain the profile line presented in this figure, the extent of this region is also shown (in pink) on figure \ref{fig:interferogram}.
	}
	\label{fig:density}
\end{figure}

Figure \ref{fig:density} shows a profile of line integrated electron density, extracted from the region of the interferogram indicated in figure \ref{fig:interferogram}. The figure also shows a (black-dashed) linear fit to the experimental data. The gradient of this trend line is $1.1 \pm 0.1 \times 10^{20} \; \si{\per\centi\meter\squared}/\mathrm{cm}$. For reference, the acceptance angle of the interferometer is $\sim \SI{20}{\milli\radian}$ which corresponds to a maximum measurable gradient of $\sim 3\times 10^{20} \;\si{\per\centi\meter\squared}/\mathrm{cm}$. From the experiments carried out so far, it seems as if there is some variation in this profile with time, and that it becomes both shallower and less well approximated by a linear trend when the distance between the target and the pinch is increased. That said, more experimental work is required to quantify this statement.              

\begin{figure}
	\centering
	\includegraphics[width=\columnwidth]{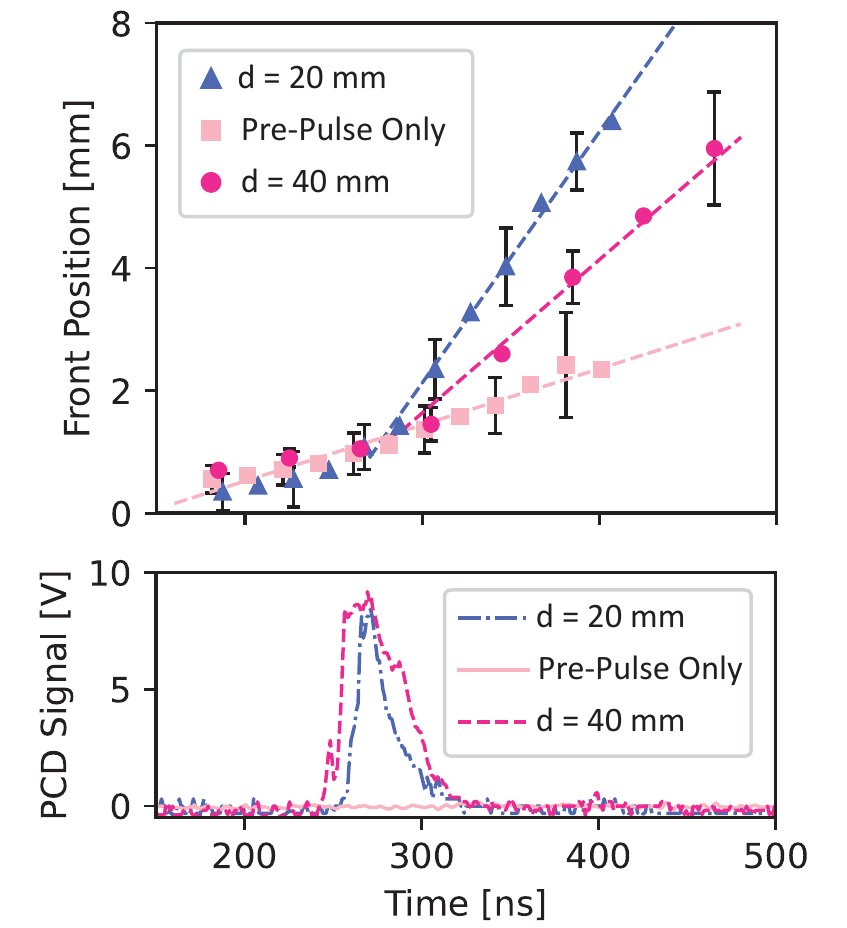}
	\caption{\emph{Top: }Position of ablation front measured from optical self emission images, captured with a fast-framing camera in three separate experiments. \emph{Bottom: }Signal from a photo-conducting diamond (PCD), filtered with \SI{20}{\micro\meter} of Beryllium foil ($\varepsilon \gtrsim \SI{700}{\electronvolt}$), and capturing emission from the Z-Pinch in the same three experiments.} 
	\label{fig:self_emission}    
\end{figure}

To obtain information about the dynamics of experiments, the MAGPIE diagnostic suite incorporates an optical fast-framing camera, fielded along the same line of sight as the laser interferometer. This diagnostic enables multiple, time gated, images of plasma self-emission to be captured in the same experiment. Data extracted from these images is shown in figure \ref{fig:self_emission}. The positions indicated in the figure correspond to the foot of the emission profile in each of the  images. The plot shows front trajectories from three separate experiments. For the data labelled with blue triangles, the separation between the pinch and the silicon target (denoted $d$) was \SI{20}{\milli\meter}. The data labelled with light pink squares used the same target separation however the Z-pinch  was `over-massed' (see \cite{HarveyThompson2009}) to prevent implosion, so the experiment was only driven with precursor/pre-pulse radiation. The final experiment (labelled with dark pink circles) used an imploding array and a target separation of $d=\SI{40}{\milli\meter}$.
      
\begin{figure}
	\centering
	\includegraphics[width=\columnwidth]{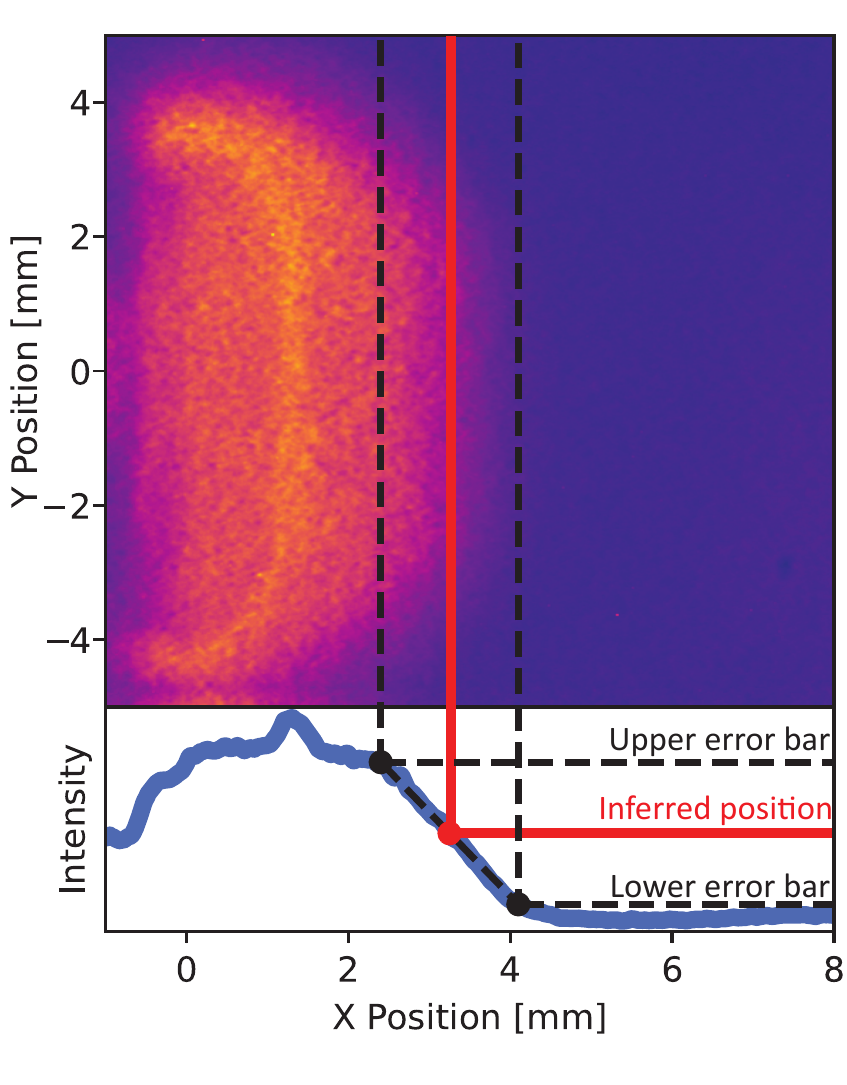}
	\caption{\emph{Top:} An example self emission image, captured \SI{385}{\nano\second} after current start in the experiment with a target separation of \SI{40}{\milli\meter}. \emph{Bottom:} An intensity profile, along the line $Z=\SI{0}{\milli\meter}$, with the inferred front position and the associated confidence interval marked upon it.}  
	\label{fig:self_emission_im}
\end{figure}

Figure \ref{fig:self_emission_im} shows an example self emission image from the experiment with a target separation of \SI{40}{\milli\meter}, along with a profile line along $Z=\SI{0}{\milli\meter}$. This figure is intended to illustrate the process used to extract the front positions that are shown in figure \ref{fig:self_emission}. This process constituted of manually picking the start and end of the rise in image intensity associated with the ablation front. The point halfway between the start and end was interpreted as the front position. The distance between the start and end was interpreted as the associated error. There are clearly other definitions of front position which one could use in analysing this data -- the one we used was chosen because it was relatively insensitive to the differing intensity response of the individual self-emission frames in the series.       

A comparison between self-emission images and interferometry data (captured along the same line of sight) revealed that the front position as measured in self-emission corresponded to a smaller $X$ value than the front position which was measured using interferometry. For example, in the shot where $d=\SI{20}{\milli\meter}$, an interferogram was captured at \SI{305}{\nano\second}. In this interferogram the leading edge of density profile was \SI{3.2}{\milli\meter} from the target. At this time, the front position in self-emission was at $X=\SI{2.7}{\milli\meter}$. The areal density measured from the interferogram at this $X$ position was $5\times10^{18}\;\si{\per\centi\meter\squared}$. All of this suggests that, in general, it is not clear how features in self-emission can be directly related to plasma conditions. That said, these front trajectories do provide a useful insight into the dynamics of the experiments.

Figure \ref{fig:self_emission} also shows the trace from a Beryllium filtered photo-conducting diamond (PCD) detector which was fielded in all three experiments. This diagnostic is sensitive to X-Rays above around \SI{700}{\electronvolt}, and these are only produced after implosion. By comparing the two plots in the figure, it is evident that the velocity of the ablation front increases at the time of implosion, in both the experiments where the array imploded. After the Z-Pinch X-rays switch off, the velocity of the front remains constant for the duration of the period which was diagnosed. 

In the experiment where $d=d_1=\SI{20}{\milli\meter}$ the measured front velocity, $V_{1} = 4.3 \pm 0.1\times 10 ^6\; \si{\centi\meter\per\second}$. In the experiment where $d=d_2=\SI{40}{\milli\meter}$, the velocity was reduced to $V_{2}= 3.3 \pm 0.1\times 10 ^6\; \si{\centi\meter\per\second}$. This difference is broadly consistent with the same fraction of deposited X-Ray energy being converted into flow kinetic energy for both target positions. To a course approximation, the pinch is a line-source of radiation and so we expect the intensity on target to decrease as $1/d$. This means that the energy ratio deposited for the two target positions is given by $d_2/d_1=2$. The ratio of the flow kinetic energies at the two positions may be approximated by $(V_{2} / V_{1})^2 = 1.7 $. These two values are consistent considering the high error in the use of self-emission images to diagnose plasma flow velocities.         

Prior to implosion, all three experiments produced fronts with a similar trajectory, and a measured velocity of $1.3\pm0.1\times 10^6 \; \si{\centi\meter\per\second}$. At early times for all three experiments, and for the entire duration of the pre-pulse only experiment, the driving radiation field is dominated by emission from the pre-pulse alone. As discussed in section \ref{sec:setup}, the pre-pulse radiation is emitted by the wires in the array and by the inflowing plasma from the wires. Therefore, the pre-pulse is probably best thought of as an extended radiation source. By contrast, emission during the main implosion is produced by the stagnated plasma on axis so behaves as a line source. The difference in sensitivity to standoff distance, which is seen in the data, can therefore be attributed to the difference in the source-size for the precursor and that of the fully imploded pinch. This is because an extended source produces an intensity scaling which is less sensitive to standoff distance than a line source.
 
The key points to take away from the discussion in this section are that the plasma profiles produced in silicon ablation experiments have a simple (quasi-1D) morphology, and that the experimental dynamics are sensitive to the time-history of the driving radiation pulse. In the next section we will compare these measurements to results from radiative magnetohydrodynamics simulations, performed with reduced models for the external radiation and magnetic fields. This comparison suggests that the presence of an ambient magnetic field plays an important role in determining the electron density profile in ablated silicon plasmas.        

\section{\label{sec:sims}Radiation magnetohydrodynamics simulations of ablated silicon plasmas}
The radiation magnetohydrodynamics (R-MHD) code Chimera \cite{Chittenden2016,McGlinchey2018}. was used to perform 2D simulations investigating the ablated plasma dynamics and effect of magnetic field in the silicon ablation experiments. Chimera is an Eulerian R-MHD code with $P_{1/3}$ multigroup radiation transport using opacity data provided by the atomic code SpK \cite{Niasse2011}; flux limited Spitzer-Harm thermal transport; extended MHD capabilities; and equation of state tables from FEoS \cite{Faik2018}.

For the work we present here, idealized radiation and magnetic field sources were used to drive a two dimensional $X-Z$ simulation with magnetic field along $Y$. The temporal variation, and spectral character of the radiation source matched the plot shown in figure \ref{fig:ideal_fields}. The intensity of the radiation was scaled until the expansion profile and front velocity matched the experimental results presented in the previous section. This corresponded to a peak brightness temperature on target of approximately \SI{10}{\electronvolt}. The size of the X-Ray source was not considered in the simulation as it was assumed that the target subtended a sufficiently small solid angle for the level of illumination to be uniform. This assumption was justified from the results of the Visrad simulations which we briefly discussed in section \ref{sec:setup}. The previous publication referenced in section \ref{sec:setup} suggests that level of irradiation from the trailing mass left during the implosion of the wire array was negligible compared to the irradiation from the stagnated Z-Pinch on axis. The contribution from the trailing mass was therefore not considered in the simulation.      

\begin{figure}
	\centering
	\includegraphics[width=\columnwidth]{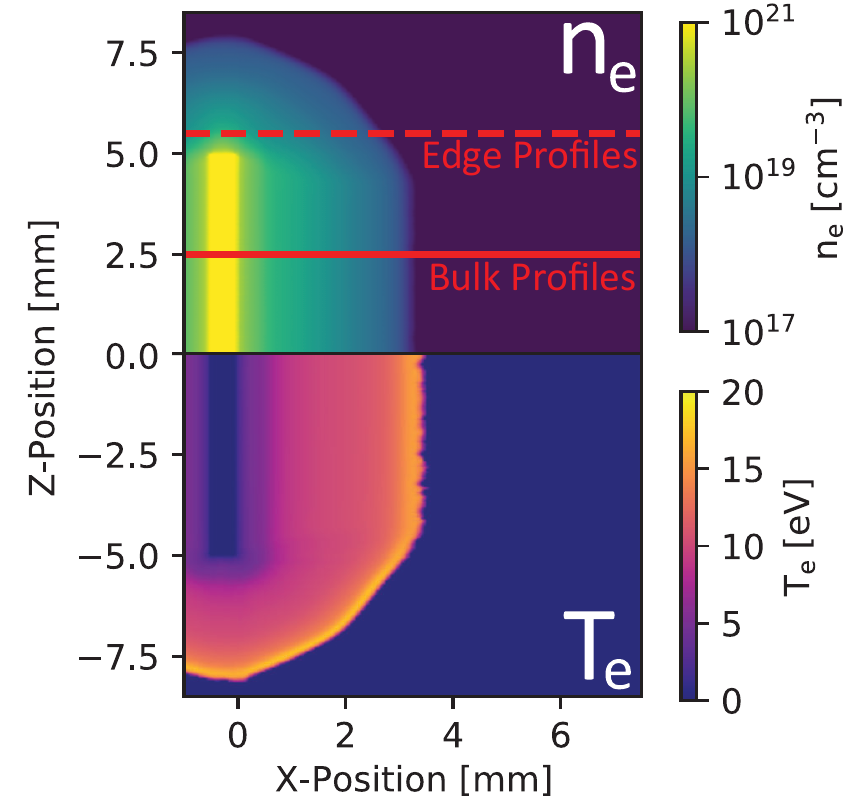}
	\caption{
		\emph{Top:} Electron density map obtained from a 2D R-MHD simulation performed with the code Chimera. \emph{Bottom:} Electron temperature map from a 2D R-MHD simulation performed with the code Chimera.
	}
	\label{fig:sim_heatmap}
\end{figure}

Spatial maps of electron density and electron temperature from a Chimera simulation are shown in figure \ref{fig:sim_heatmap}. In general, a comparison with the interferometry data presented in figure \ref{fig:interferogram}, shows that the morphology of ablated plasma is in good agreement with that seen in experiments. Close inspection of the temperature map shows a slight corrugation at the leading edge of the expanding plasma. These corrugations grew from grid-level numerical noise in the simulation and no such feature is seen in the electron density map. Analysis of the various contributions to total pressure in the simulations revealed that this corrugation was imprinted on the temperature at a very early time: Subsequently the plasma and magnetic pressures relaxed such that the edge of the expanding plasma was close to isobaric with the vacuum magnetic field, preventing instability growth.         

\begin{figure}
	\centering
	\includegraphics[width=\columnwidth]{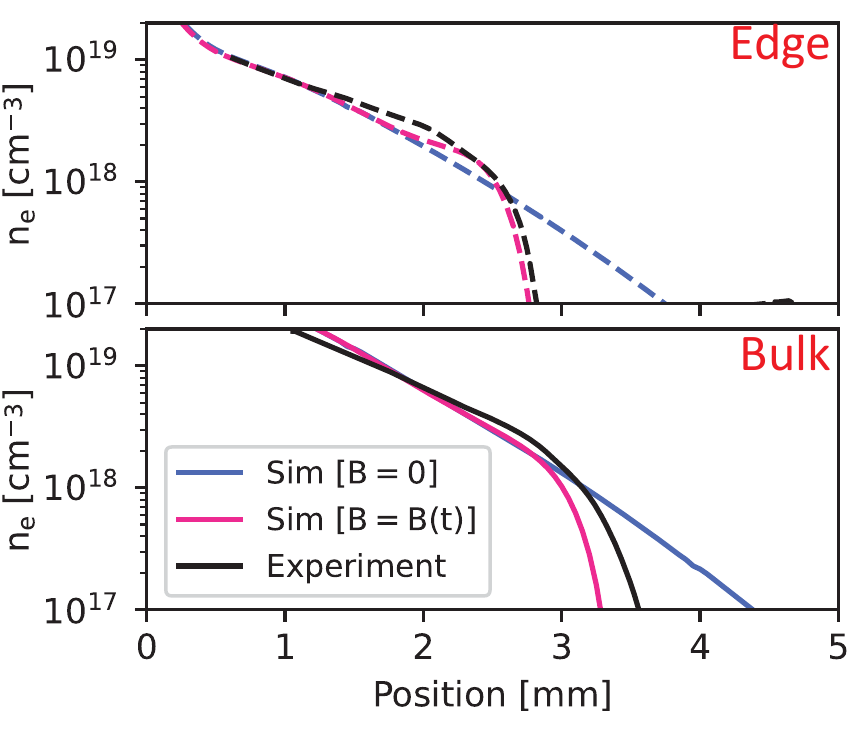}
	\caption{
		Comparison of electron density profiles between experimental data and R-MHD simulations performed with / without an applied magnetic field. The range of the vertical axes were chosen to roughly match the dynamic range of the experimental data. The results show that the B-Field tamped the expansion of the ablated silicon plasma. \emph{Top:} Profiles for plasma conditions above the top edge of the silicon target. \emph{Bottom:} Profiles for plasma conditions in the bulk (middle) of the expansion.
	}
	\label{fig:sim_density}
\end{figure}

Figure \ref{fig:sim_density} shows profiles in electron density taken from two different $Z$ positions. These two positions are indicated by horizontal lines in figure \ref{fig:sim_heatmap}. The first location was close to the middle of the target, and is labelled `bulk'. The second location was \SI{0.5}{\milli\meter} above the top of the target and is labelled `edge'. Three separate profiles are plotted for each $Z$ location. The first of these (rendered in black) is a profile obtained from interferometry data. The other two profiles are from simulations performed with and without an applied magnetic field (rendered in pink and blue respectively). Note that the extent of the vertical axis in the figure roughly corresponds to the dynamic range of the interferometer which was used to obtain the experimental profiles. This means that the visible trend in the experimentally measured profiles is not affected by diagnostic saturation. 

The lower limit on the response of the interferometer was set by the smallest fringe shift which could be detected. A density of $10^{17} \; \si{\per\centi\meter\squared}$ corresponds to a shift of 0.25 fringes, approximately the limit on the accuracy of the fringe tracing process. The upper limit on the diagnostic response was set when the fringe width approached size of the pixels on the detector, for these experiments at a density of around $10^{19}\; \si{\per\centi\meter\squared}$.    
 
A comparison of the three profiles shown in figure \ref{fig:sim_density} demonstrates that the presence of an external magnetic field in the simulations acted to tamp the expansion of the ablated silicon plasma. Furthermore the plot shows that the presence of an external magnetic field in the simulations was required in order for them to match the experimental data.

To summarise the discussion in this section, the Chimera R-MHD simulations we presented were able to accurately reproduce the density profiles and overall morphology seen in silicon ablation experiments. Furthermore, the simulations demonstrated that the presence of an external magnetic field played an important role in determining the form of electron density profiles.     
 
\section{\label{sec:thomson}Diagnosis of temperature and velocity profiles with optical Thomson scattering}
Thomson scattering is a diagnostic which involves shining a focussing probing laser beam through a plasma and recording the spectrum of light which is Thomson scattered by the electrons in the plasma \cite{Sheffield2010}. The form of the spectrum is sensitive to various plasma parameters, but for the work we present here the diagnostic was used to measure profiles of temperature, velocity, and average degree of ionisation.

\begin{figure}
	\centering
	\includegraphics[width=\columnwidth]{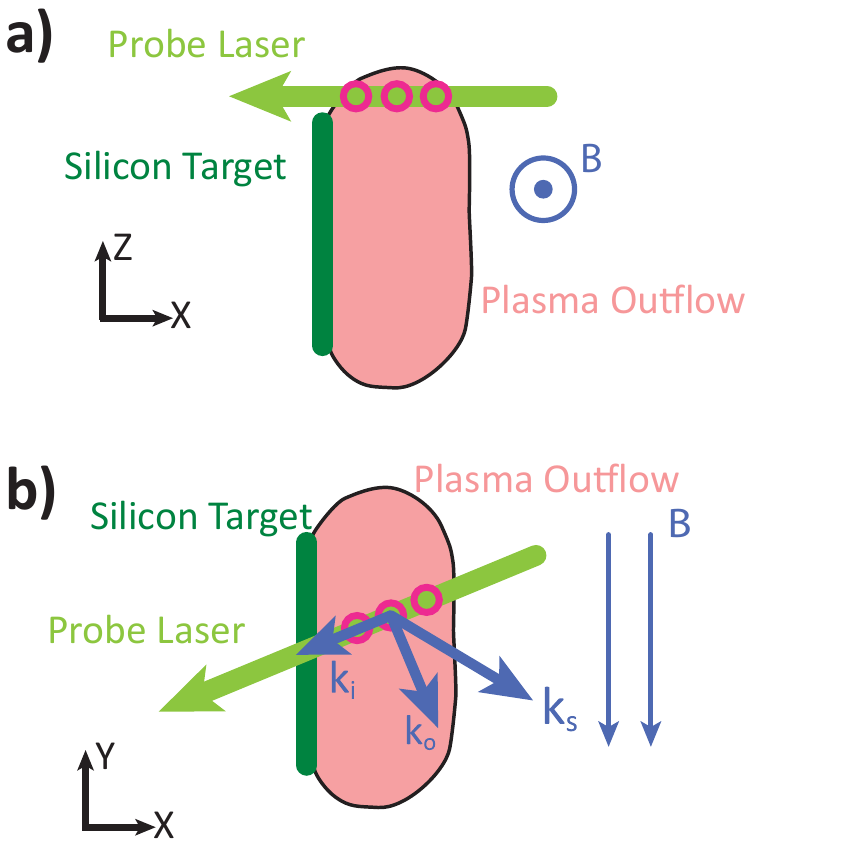}
	\caption{
		Diagrams of the Thomson scattering geometry which was used in the work discussed here. The Thomson probe was directed above the target, in order to avoid dumping its energy into the experimental volume. a) Shows a \emph{`side-on'} view of the experiment, where structure is resolved in the $X-Z$ plane. b) Shows an \emph{`end-on'} view of the experiment, where structure is resolved in the $X-Y$ plane.
	}
	\label{fig:ts_gemometry} 
\end{figure}

Thomson scattering data was obtained for the setup which used a \SI{40}{\milli\meter} target separation. The diagnostic setup enabled the measurement of scattering spectra from a number of spatially localised volumes. Figure \ref{fig:ts_gemometry} shows the scattering geometry which was used. In the diagram,  $\mathbf{k_i}$ is the wave-vector of the incoming (probing) laser; $\mathbf{k_o}$ is the wave-vector of the scattered light; and $\mathbf{k_s}$ is the wave-vector of a plasma mode. The three vectors are connected by the momentum conservation rule: 
\begin{equation}
\label{eq:ts_momentum}
\mathbf{k_s} = \mathbf{k_o} - \mathbf{k_i}.
\end{equation}
As shown in figure \ref{fig:ts_gemometry}, for this experiment, $\mathbf{k_s}$ lay in the $X-Y$ plane (parallel to the magnetic field driven by the pinch). The angle between the scattering vector and target normal was \SI{22.5}{\deg}. The $Z$ position of the probing laser was set so it passed just above the silicon target. This was done to allow the laser pulse to exit the experimental volume, rather than dumping its energy into the target and perturbing the plasma conditions.

The probing laser  used frequency doubled Nd:YAG ($\lambda=\SI{532}{\nano\meter}$) in a \SI{2}{\joule}, \SI{4}{\nano\second} pulse. The scattered light was imaged onto a linear array of 14 optical fibers, and so data was obtained from 14 (spatially separated) volumes. Time-gating was provided by an intensified CCD and the exposure time was \SI{4}{\nano\second}. A more detailed description of the diagnostic is given in \cite{Suttle2021}. 

The ion acoustic feature was diagnosed, and so scattering profiles (in general) were sensitive to the projection of plasma velocity onto the scattering vector  ($\mathbf{\hat{k_s} \cdot V}$); the product of electron temperature and average ionisation ($T_e \times \overline{Z}$); ion temperature ($T_i$); and electron density ($n_e$) \cite{Sheffield2010}.

The way scattering spectra respond to changes in $n_e$ and $T_i$ is broadly similar, and so measurement of the ion-acoustic feature alone does not provide an independent diagnosis of these two parameters. For the fits presented here $n_e$ was constrained from interferometry data, captured \SI{10}{\nano\second} before the Thomson scattering data. This timescale is very short compared to the characteristic timescales of the experiment (see for example figure \ref{fig:self_emission}).

An important detail is that we also \emph{enforced temperature equilibration} in our fits (i.e. stipulated that $T_e = T_i = T$) and \emph{treated $\overline{Z}$ as an independent fitting parameter}. This was justified because the ion electron temperature equilibration time was short compared to timescales associated with the hydrodynamics ($\tau_{ei}\lesssim\SI{1}{\nano\second}$) and so we do not expect significant ion electron temperature separation.   

\begin{figure}
	\centering
	\includegraphics[width=\columnwidth]{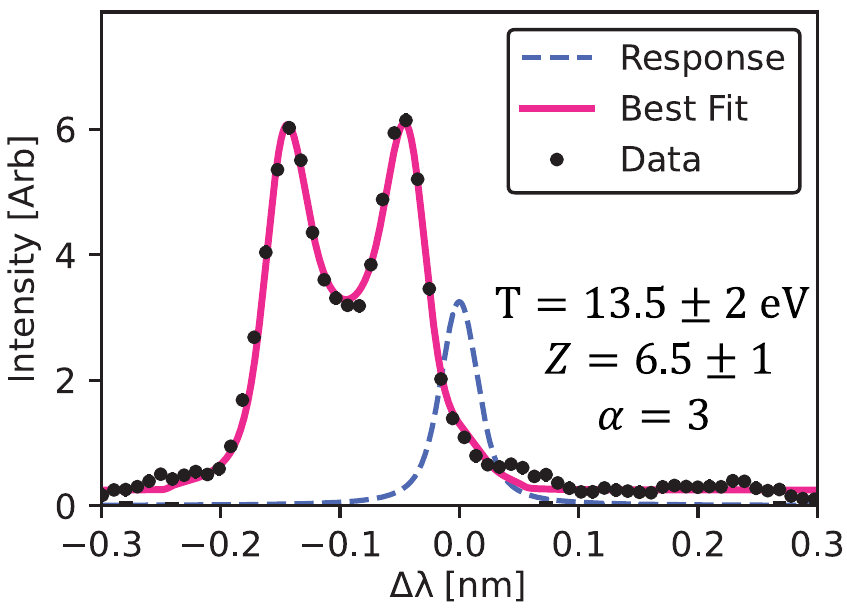}
	\caption{
		A typical spectrum of Thomson scattered light, taken from the volume at $X=\SI{1.8}{\milli\meter}$. The spectrum is associated with ion-acoustic fluctuations, and sits well within the collective scattering regime ($1/[k_s \lambda_D ] = \alpha = 3.0$).
	}
	\label{fig:thomson_spectrum}   
\end{figure}
A typical Thomson scattering spectrum for the plasma conditions in this experiment is presented in figure \ref{fig:thomson_spectrum}. This shows that Landau damping was sufficiently small to allow for the presence of two clear ion acoustic peaks in our data. These two peaks are formed by ion acoustic fluctuations with wave-vectors parallel and antiparallel to $\mathbf{k_s}$. An important detail to note is that the response (i.e. resolution) of the spectrometer was comparable to the apparent width of the ion acoustic peaks. Thus, the calculated scattering spectra were convolved with the response in order to correct for instrumental broadening. 

\begin{figure}
	\centering
	\includegraphics[width=\columnwidth]{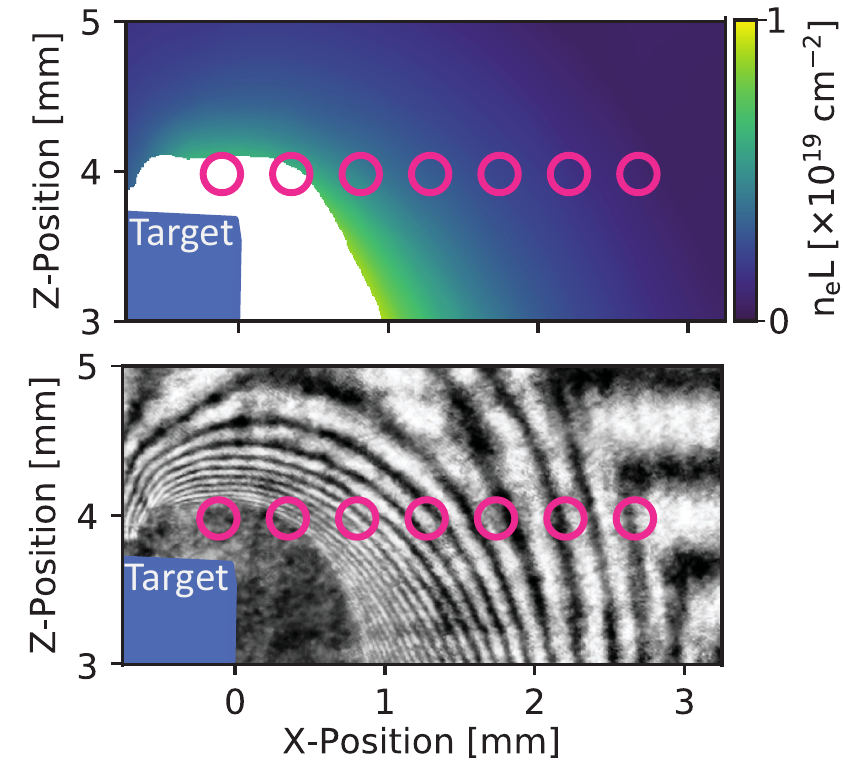}
	\caption{
		Position of Thomson scattering volumes overlaid on measured electron density data. \emph{Above:} Volumes are overlaid on processed interferometry data. \emph{Below:} Volumes are overlaid on the raw interferogram which was used to obtain density profiles.   	
	}
	\label{fig:thomson_interferometry}
\end{figure}
The interferometry data which was used to constrain Thomson fits is shown in figure \ref{fig:thomson_interferometry}. The positions of scattering volumes are overlaid as pink circles in the figure. The size of these circles is also equivalent to the size of scattering volumes. The figure shows both raw interferometry data and a processed electron density map. Details of the procedure which was used to extract the processed map from a raw interferogram are provided in \cite{Hare2019}. For the two points which are nearest to the target, the interferometry probe laser was totally absorbed via inverse-Bremsstrahlung and so an interferometry signal was not observed. To obtain an indicative value of $n_e$ for these two points, a linear fit to values of $n_e$ in the region $1 < X <\SI{2}{\milli\meter}$ was performed and the result was extrapolated over the $-1 < X < \SI{1}{\milli\meter}$ interval. As the ion acoustic spectrum has a relatively weak dependence on electron density, the error introduced by this extrapolation is likely to be negligible.   

\begin{figure}
	\centering
	\includegraphics[width=\columnwidth]{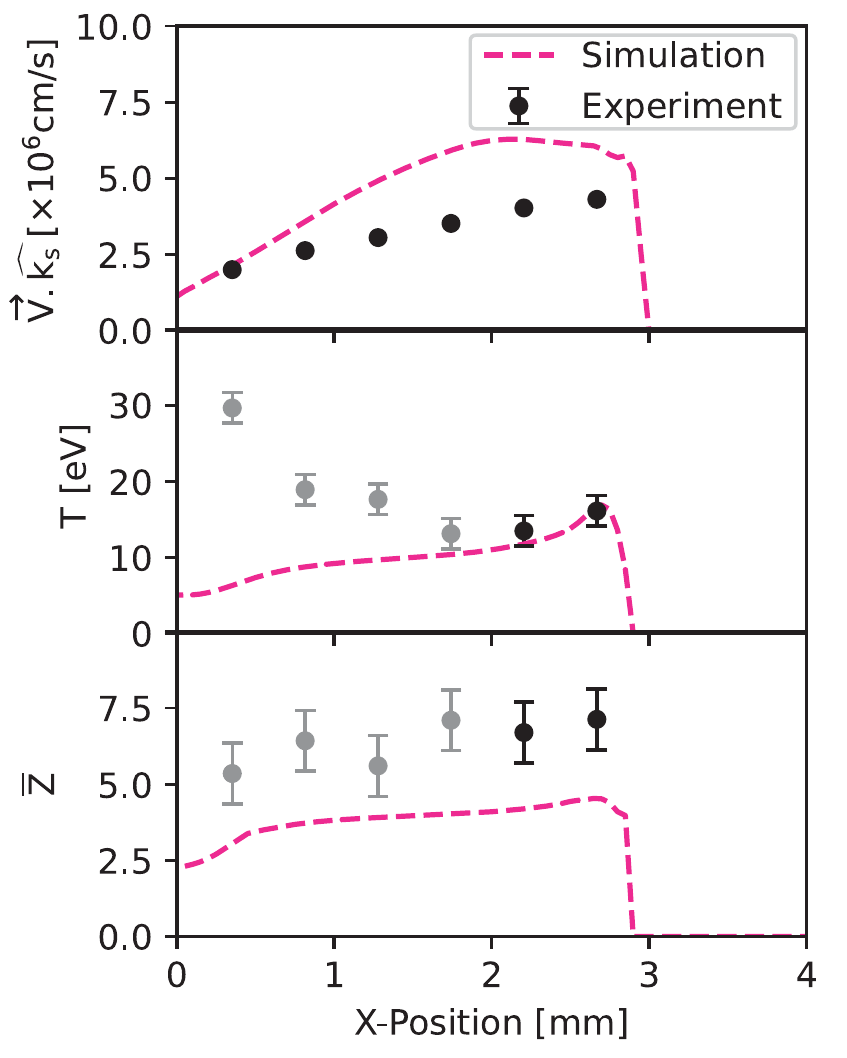}
	\caption{
		Profiles of $ \mathbf{V} \cdot \hat{\mathbf{k_s}} $, $T_e$, and $\overline{Z}$ from Thomson scattering compared with profiles from a Chimera simulation. Points plotted in grey represent Thomson volumes in which probe heating significantly effected plasma conditions.
	}
	\label{fig:thomson_profile}    
\end{figure}
Figure \ref{fig:thomson_profile} shows spatial profiles in velocity, temperature, and average degree of ionisation which were measured from Thomson scattering, compared against results from a Chimera R-MHD simulation. The Chimera results were taken from a position above the top of the simulated target, in order to account for edge effects introduced by the Z position of the Thomson probe in the experiment.
 
Looking first at the velocity profiles, in figure \ref{fig:thomson_profile}, note that both the experimental and simulated profiles considered the velocity component parallel to the scattering vector (i.e. $\mathbf{V} \cdot \hat{\mathbf{k_s}}$) and so may be directly compared. Overall, the shape of the trend seen in both simulation and experiment is broadly comparable however there is a systematic offset between the two datasets. The reason for this remains under investigation. We have investigated the affect of dimensionality by performing 3D simulations. The velocity profile observed in 3D is unchanged from the 2D result shown here.

Another possibility is that the velocity profile seen in experiment may be sensitive to aspects of the time-history or spectral character of the radiation drive which were not captured in the reduced model of the external radiation field used for the simulations presented here. There is an ongoing effort to better capture the radiation drive using spectra obtained by post-processing integrated Gorgon simulations of the Z-Pinch with the atomic code SpK. The data shown in figure \ref{fig:spk_spec} are preliminary results from these simulations.           
  
Looking now at the temperature data, we see that there is good agreement between simulation and experiment for the two outermost Thomson scattering volumes in figure \ref{fig:thomson_profile}. For the remainder of the volumes (which are coloured grey in the plot), the trend seen in the data diverges from the one seen in the simulation. We believe that this can be attributed to the Thomson probe heating the plasma in the experiment, thereby perturbing the measured temperature.

\begin{figure}
	\centering
	\includegraphics[width=\columnwidth]{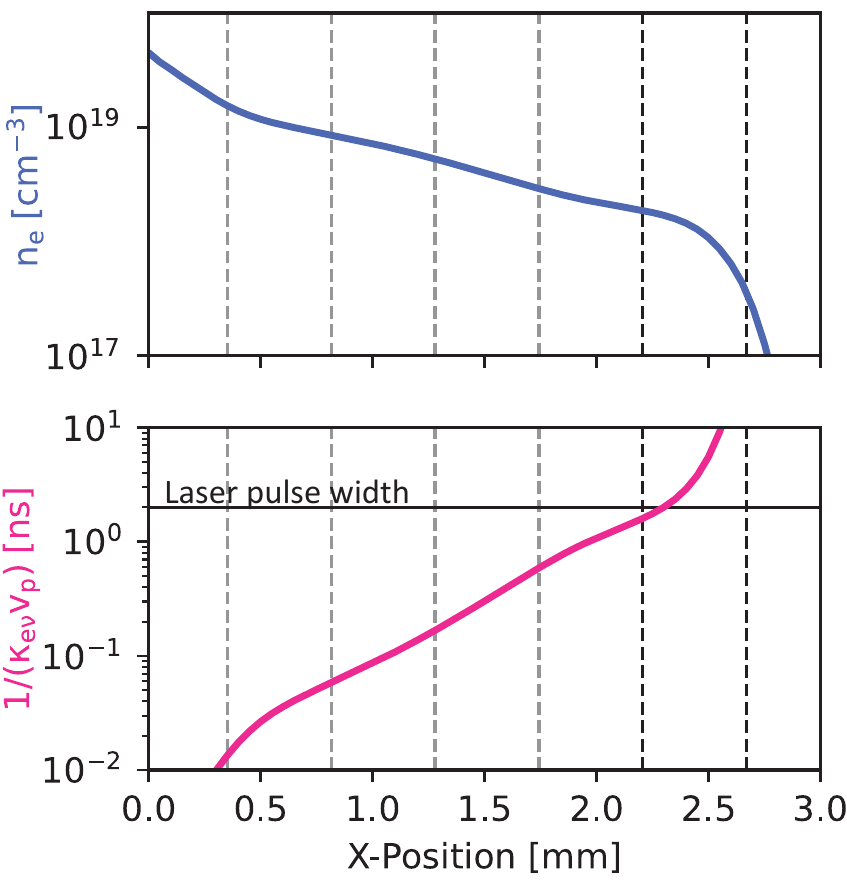}
	\caption{
	\emph{Above:} Profile of electron density from a Chimera simulation. \emph{Below:} Profile of the timescale associated with inverse-Bremsstrahlung heating calculated from simulated plasma parameters. The horizontal line (in the lower plot) corresponds to the temporal-width of the probing laser pulse. The vertical dashed lines (in both plots) indicate the positions of the Thomson scattering volumes.
	}
	\label{fig:kappa_simprofile}   	   
\end{figure}
This picture is broadly consistent with the results seen in simulation, a point which is demonstrated in figure \ref{fig:kappa_simprofile}. This plot shows simulated plasma density (above, in blue) and an estimate of the timescale assosiated with inverse-Bremsstrahlung heating (below, in pink). The horizontal solid line in the lower plot denotes the full width half maximum (FWHM) of the probe laser pulse (in nanoseconds) and the vertical dashed lines indicate the positions of the Thomson scattering volumes in the experiment. 

The estimate for the timescale associated with heating is given by 
\begin{equation}
\tau_{e\nu} = \frac{1}{\kappa_{e\nu}\mathrm{v_p}},
\end{equation} 
where $\kappa_{e\nu}$ is the inverse-Bremsstrahlung absorption coefficient (calculated using the expression given in \cite{Richardson2019}, with simulated plasma parameters) and $\mathrm v_p$ is the phase speed of the probing laser light. 

The quantity $\tau_{e\nu}$ represents an approximation to the timescale over which plasma temperature might increase by an e-folding. This model neglects the fact that the heat deposited by the probe will be transported to its surroundings; that the laser light will be depleted by absorption; and that as the plasma is heated it becomes less absorbative, so in reality it is likely an overestimate. That said, the picture this very coarse treatment paints is in broad agreement with the one seen in our experimental data: The heating timescales for the first two Thomson volumes are comparable to, or greater than, the FWHM of the probe laser pulse so we do not expect them to be significantly heated before the laser turns off. By contrast, the remainder of the Thomson volumes are expected to be heated on a timescale much shorter than the interval for which the probe is switched on. Returning to figure \ref{fig:thomson_profile}, we see that this is  consistent with the difference seen between the simulation and the Thomson scattering data.

In-spite of the fact that the Thomson probe appears to be perturbing plasma conditions, we still believe that the velocity profile obtained by the diagnostic is representative of the unperturbed plasma. Our evidence for this is largely empirical: Experiments have been performed on the COBRA pulsed power generator (Cornell Laboratory of Plasma Studies) with comparable plasma conditions and a temporally streaked Thomson scattering diagnostic. In these experiments a significant perturbation to the measured velocity was not observed, even when probe heating was quite dramatic \cite{Banasek2021}.

Returning to figure \ref{fig:thomson_profile}, note that the average degree of ionisation in the Thomson scattering data is significantly higher than the result seen in the simulation, even for the volumes in which measured temperatures are consistent with the simulation. We emphasise that the uncertainty in the experimental data is too large to have confidence in this result -- it is possible that the threshold for heating disrupting the measured charge state is more stringent than the threshold for the measurement of plasma temperature. That said, if these initial results are confirmed by further systematic measurements, then this would be an interesting result. One possibility is that the presence of the external radiation field (produced by the wire array Z-Pinch) might be affecting the charge state distribution in the ablated silicon plasma. 

To investigate this possibility we have performed a series of simulations with the 1D Lagrangian Hydrodynamics code Helios-CR \cite{MacFarlane2006}. This code does not treat MHD effects, but includes a sophisticated inline non-equilibrium atomic kinetics and radiation transport capability. The model which the code uses couples the X-Ray flux to the atomic kinetics of the plasma with radiation-dependent photoionization and photoexcitation rates. Our preliminary findings from these simulations suggest that the driving radiation field indeed plays a role in determining the charge states present in the ablated plasma flows. When this work reaches a greater degree of maturity we intend to report on these findings in a follow-up publication.

To summarise discussion in this section, we reported on Thomson scattering results in ablated silicon plasmas. We found that the trend seen in plasma velocity measurements was broadly consistent with results from R-MHD simulations. Experimental measurements of temperature were either consistent with the results from simulations or were compromised by Thomson probe heating. Preliminary results indicate that the average charge state seen in experiment was higher than might be expected from Chimera simulations however this finding carries a very high degree of uncertainty. 
     
\section{\label{sec:absorbtion}Diagnosis of the Inverse Bremsstrahlung absorption coefficient}
Motivated by a desire to verify simulated the temperature profiles with an independent diagnostic, we performed experiments in which the optical absorption via inverse Bremsstrahlung was diagnosed experimentally. The functional form of the inverse Bremsstrahlung absorption coefficient is
\begin{equation}
\kappa_{e\nu} [\si{\per\centi\meter}] = \frac{3.1\times 10^{-7} \overline{Z} (n_e [\si{\per\centi\meter\cubed}])^2 \ln\Lambda}{(T_e[\si{\electronvolt}])^{3/2} (\omega [\si{\radian\per\second}])^2 \sqrt{1-\left(\frac{\omega}{\omega_p}\right)^2}},
\label{eq:kappa} 
\end{equation}            
where $\ln\Lambda$ is the Colomb logarithm; $\omega$ is the angular frequency of the probing laser; and $\omega_p$ is the electron plasma frequency \cite{Richardson2019}. For the work presented here, we assumed the form of $\ln\Lambda$ given in \cite{Johnston1973}. 

Looking at equation \ref{eq:kappa}, it can be seen that the primary dependencies in $\kappa_{e\nu}$ are $n_e$ and $T_e$, as $\ln\Lambda$ only depends weakly on plasma parameters. Note that, for these conditions, $\omega_p << \omega$ and so the variation introduced by the $\omega/\omega_p$ term is negligible. The results presented above demonstrate that $n_e$ is well matched between experiment and simulation so a comparison between the value of $\kappa_{e\nu}$ measured in experiment and the value calculated from simulated plasma parameters can be used to verify the plasma temperature profile obtained in the simulations.

In order to measure $\kappa_{e\nu}$ we obtained two laser shadowgrams of the setup. One shadowgram was a background, obtained before the experiment was performed, and so no plasma was present in this image. A second shadowgram was obtained during the shot, and so an outflowing silicon plasma was present. In the following discussion, we denote the intensity profile in the background image $I_0(x,y)$, and the profile in the shot image $I(x,y)$. To perform this measurement, it was necessary to use a sufficiently stable laser beam -- the result we report was obtained using the Cerberus laser facility, which is sufficiently stable. The use of this facility to make imaging Faraday rotation measurements, a diagnostic technique which bears some similarity to the one we describe here, is presented in reference \cite{Swadling2014}.          
                    
\begin{figure}
	\centering
	\includegraphics[width=\columnwidth]{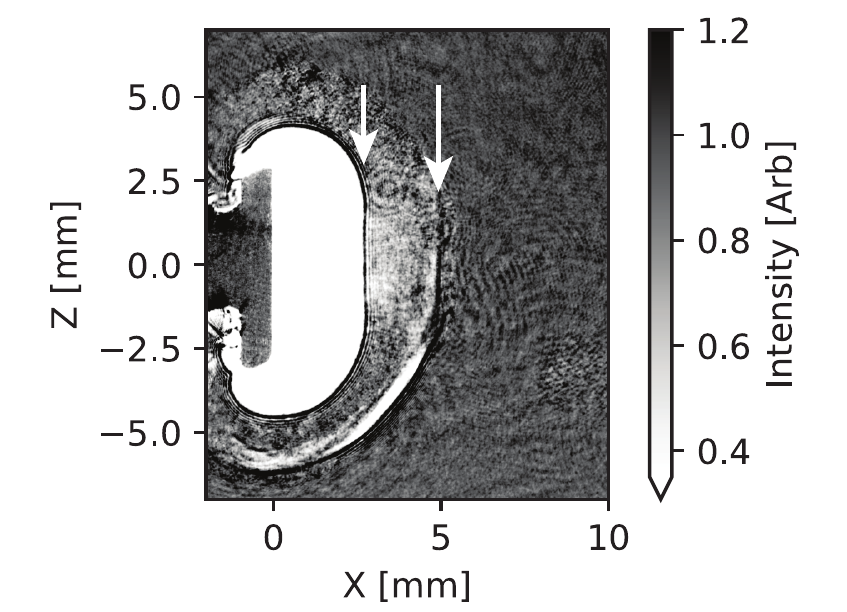}
	\caption{
		A shadowgraphy image of a silicon absorption experiment. Arrows indicate caustics, caused by strong shadowgraphy effects. 
	}
	\label{fig:shadowgram} 
\end{figure}
The image in figure \ref{fig:shadowgram} is normalised shadowgraphy data, obtained by calculating $I(x,y)/I_0(x,y)$. In the regions of the image indicated with white arrows, strong shadowgraphy effects are present, and the image carries information about density gradients. In the region of ablated plasma bounded by the two arrows, density gradients are relatively weak and so the reduction in intensity seen in the data may be attributed to laser absorption rather than shadowgraphy. 
It is apparent that the value of $\kappa_{e\nu}$ can be obtained within this region using the equation   
\begin{equation}
\kappa_{e\nu} = \frac{-\ln(I(x,y)/I_0(x,y))}{L_y},
\end{equation}
where $L_y$ is the extent of the object in the $y$ direction (measured using orthogonal laser probing). 
 
\begin{figure}
	\centering
	\includegraphics[width=\columnwidth]{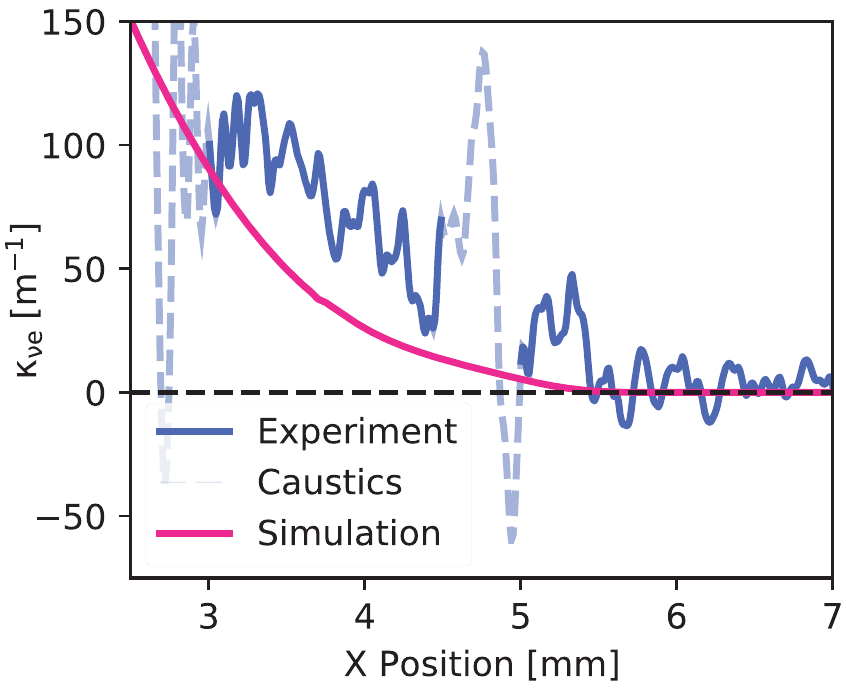}
	\caption{
		Comparison between the profile of measured inverse-Bremsstrahlung absorption coefficient and the profile predicted from a simulation.
	}
\label{fig:kappa_expprofile}   
\end{figure}
Figure \ref{fig:kappa_expprofile} shows the profile of $\kappa_{e\nu}$ measured in an experiment compared against the one calculated with equation \ref{eq:kappa}, using simulated plasma parameters. The regions where experimental data are affected strongly by shadowgraphy are greyed out in the graph. 

In the first instance, it is apparent that the general shape of the two curves are in broad agreement -- this suggests that the temperature profile obtained in the simulation is representative of the physical system: A good sign. That said, there is a slight systematic offset between the two curves. One possible reason for this is an error in the value of $L_y$ assumed in the analysis. Another possibility is that some aspect of the simulated plasma parameters (for example the degree of ionisation) may fail to reproduce the conditions of the experiment. A final (more interesting) possibility is that the discrepancy can be attributed to an error in the determination of the Coulomb logarithm. This is not unexpected as the data we present is in a regime where $\ln\Lambda$ is approaching two. It has been reported that the theory is prone to error in this regime \cite{Skupsky1987}. In the future, if we are able to find a reliable method to diagnose temperature and the charge state distribution, then a compelling use for the platform we describe is the direct measurement of inverse Bremsstrahlung absorption coefficient at different optical wavelengths.    

\section{\label{sec:conclusion}Conclusions}
In this paper we presented some initial results from a novel experimental platform which used the radiation pulse from a \SI{1.4}{\mega\ampere} wire-array Z-Pinch to drive ablation from a solid silicon target. The plasma produced by the radiative ablation process expanded into an ambient magnetic field, supported by the current flowing through the Z-Pinch. 

Experiments were diagnosed with laser interferometry, optical fast frame imaging, Thomson scattering, and laser shadowgraphy. Simulations of the experiment were performed using the radiative-magnetohydrodynamics code Chimera \cite{Chittenden2016,McGlinchey2018} and the radiative-hydrodynamics code Helios-CR \cite{MacFarlane2006}. 

Initial experimental results indicate that the plasma flows produced in experiments have a simple (quasi-1D) morphology, and that the experimental dynamics are sensitive to the time history of the driving radiation pulse. Results from Chimera simulations suggest that electron density profiles are modified by the presence of an external, ambient magnetic field. Thompson scattering data suggests that the degree of ionisation in the ablated plasma flows may be modified by the presence of an external radiation field from the Z-pinch and this possibility is supported by Helios-CR simulations. The laser shadowgraphy data was used to directly measure the inverse-Bremsstrahlung absorption coefficient ($\kappa_{e\nu}$), and this was compared to values of $\kappa_{e\nu}$ calculated from Chimera simulations. It was found that the measured value of the absorption coefficient was consistently higher than the value predicted from simulations. 

A key detail to take away from the work we have presented is that the radiative ablation platform produced plasma flows with a simple morphology, and well defined boundary/initial conditions. This enabled individual physical processes to be isolated and studied separately -- unusual in high energy density physics, where many of the physical processes detailed here are typically only accessible in more complex, integrated experimental schemes. 

In the near future we hope to study the penetration of magnetic flux into ablated plasma flows, using diagnostics such as Faraday rotation \cite{Swadling2014} and Thomson scattering \cite{Bruulsema2020}. We also intend to diagnose the charge state distribution of plasma flows directly by using silicon K-shell absorption spectroscopy. Additionally, if absorption spectroscopy proves to be a viable technique to diagnose the temperature and charge state distribution of the ablated plasma then we intend to use the laser absorption technique, described in section VI, to make a robust comparison with theories for inverse Bremsstrahlung absorption.    

\section*{Acknowledgements}
This work was supported by the U.S. Department of Energy (DOE) under Award Nos. DE-SC0020434 and DE-NA0003764; and by the U.S. Defence Threat Reduction Agency (DTRA) under award number HDTRA1-20-1-0001.

\section*{Data Availability}
The data that support the findings of this study are available from the corresponding author upon reasonable request.                

\bibliography{Bibliography}
\end{document}